\documentstyle[12pt]{article}
\begin{document}  
\vskip 2mm

\begin{center}
{\Large\bf The QCD observables expansion over the scheme-independent two-loop
coupling constant powers, the scheme dependence reduction}
\vspace{4mm}\\
{\large D.S.~Kourashev}\\
{\small (\today)}\\

\vspace{0.2cm}
{\it Bogoliubov Lab. Theor. Phys., JINR, Dubna, 141980, Russia}\\
{\it kourashev@mtu-net.ru}
\end{center}

\abstract{\small
The method suggested in this paper allows to express the n-th order 
renorm-group equation solutions over the powers of the two-loop solution, that can
be obtained explicitly in terms of the Lambert function.
On the one hand this expansion helps to get more reliable theoretical
predictions, on the other hand the scheme dependence problem can be
understood better. When using this method, Stevenson scheme invariant
expressions can be obtained easily, the scheme
dependence emerging from the perturbative series truncation can be
estimated and reduced. The `optimal' choice of the scale
parameter allows to have at the three-loop level
the scheme dependence magnitude corresponding to the four-loop level etc.
The new criterion, principally different from the Ra\c czka criterion,
is introduced.}

\section{Introduction}

The renorm-group equation (RGE) for the QCD coupling constant $\bar \alpha _s$ 
\begin{eqnarray}
-x\frac{\partial a}{\partial x}=a^2+b_1a^3+b_2a^4+...
\ \ \mbox{, } \ \ a=\frac {\beta_0}{4\pi }\overline{\alpha }_s
\ \mbox{, } \ \ b_n=\frac{\beta_n}{\beta_0^n}\label{rgeq}\mbox{. }
\end{eqnarray}
can be solved analytically at the one-loop level (leading order) $a^{(1)}(x)=\frac 1{\ln x}$,
the 2nd-order (next-to-leading order) RGE solution can be expressed in terms
of the Lambert function~\cite{magradze, grunberg, lambert}
\begin{eqnarray}\label{a2}
\widetilde a(x)=-\frac 1{b_1}\frac 1{1+W(z)}\ \ \mbox{, } \ \ \
z \equiv -\frac 1ex^{-\frac 1{b_1}}.
\end{eqnarray}
The W Lambert function is defined through a transcendental equation
$W(z)e^{W(z)}=z$. Here and further $x=%
\frac{Q^2}{\Lambda ^2}$, $\Lambda $ is the scale parameter defining unphysical
singularities positions. At the two-loop level unphysical cut emerges when
$Q^2<\Lambda^2$.   \par

For the 3rd order RGE the 
Pad\'e-approximated $\beta $-function~\cite{pade} can be used and the solution
can be also obtained in terms of the Lambert function. However, without Pad\'e
approximation even at the three-loop case we can't get explicitly the RGE
solution for both coupling constant and observables.\par

Another problem is observables unphysical scheme dependence
that is related with the multi-loop terms uncertainty. It originates
the theoretical uncertainty of physical results. The using of different
renorm-schemes (RS) leads to various results and there are not any reasons
to consider one of these schemes as the most preferable. The natural way
to reduce this uncertainty is to reduce the scheme dependence.\par

As an observable example the process of $e^+e^-$-annihilation into hadrons
is used in this paper. This process is rather interesting when analysing the
scheme dependence~\cite{rac&czy}.\par

The method suggested in this paper is based on the expansion of
the n-th order RGE solutions over the two-loop coupling constant~(\ref{a2})
powers. This approach proves to be very useful to analyse
the problem of observables scheme dependence, to get perturbative
coefficient invariant combinations, to estimate the theoretical
uncertainty (the new criteria was elaborated for this purpose). But the
most important consequence is the possibility of reducing this uncertainty.

\section{The expansion of
multi-loop RGE solutions over the two-loop solution powers}

We used the expansion of the n-th order RGE
solution over the powers of some known solution (obtained in the certain RS).
The attempt to obtain such an expansion over the one-loop function powers
can not be successful, because even the 2nd order RGE solution possesses
$\log \log $ singularities that couldn't be expanded over the powers of
$a^{(1)}(x)=\frac1{\ln x}$.
It can be easily understood also with the Lambert function
consideration. The two-loop coupling constant expressed as a function of
the one-loop constant
has an evident singularity in the vicinity of $a^{(1)}=0$, originating from
the Lambert function argument $z=-e^{\frac 1{b_1a^{(1)}}}.$
Let us begin with the assumption of the expansion over the two-loop function
powers existing
\begin{eqnarray}\label{expansion}a^{(n)}=\sum\limits_ik_i \widetilde{a}^i
\,.\ \
-\frac{\partial \widetilde{a}}{\partial t}=\widetilde{a}^2+b_1\widetilde{a}%
^3\,,\ \ t=\ln x, 
\end{eqnarray}
i.e. $a^{(n)}$ is n-th order RGE solution. To obtain $k_i$ values let us
consider
\begin{eqnarray}\label{koef}\sum\limits_ik_i\partial _t\widetilde{a}^i=
\sum\limits_ik_i\left(\widetilde{a}^2+b_1\widetilde{a}^3\right)=\left(
\sum\limits_ik_i \widetilde{a}^i\right) ^2+b_1\left(
\sum\limits_ik_i\widetilde{a}\right) ^3+...\end{eqnarray}
From the $\widetilde{a}(x)$ powers coefficients equality follows 
$$
k_1=1;\, k_2=?;\, k_3=k_2^2+b_1k_2+b_2;\, k_4=k_2^3+\frac 52b_1k_2^2+3b_2k_2+\frac 12b_3. 
$$
$k_2$ coefficient is arbitrary, it corresponds to the arbitrariness of the
scale parameter choice. Indeed, if we express some RGE solution through the
certain two-loop solution we should have the parameter that permits to
obtain solutions with a different scale parameter.\par

Researching of the $k_i$ asymptotic properties is complicated because
of the involved recurrent formula
$$k_i=k_{i-1} b_1 \left(-1+\frac 3 {i-2} \right)+
k_{i-2} b_2 \frac {3 b_2} {i-2}+\,.\,.\,. \mbox{(for }i>4\mbox{).}$$
Suppose $k_2=0$ (that corresponds to the natural choice, when the multi-loop
coupling constant and the two-loop constant have the same scale parameter).
So, $k_i$ values are: 
$$
k_3=b_2;\, k_4=\frac 12b_3;\, k_5=\frac 53b_2^2-\frac 16b_1b_3+\frac 13 b_4%
$$
$$
k_6=-\frac 1{12}b_1b_2^2+\frac 1{12}b_1^2b_3+2b_2b_3-\frac16b_1b_4+
\frac14b_5%
$$
$$
k_7=\frac{16}5b_2^3-\frac 45b_1b_2b_3+\frac 1{20}b_1^2b_2^2-\frac
1{20}b_1^3b_3+\frac{11}{20}b_3^2+ \frac 1{10}b_1^2b_4-\frac
3{20}b_1b_5+\frac15b_6.%
$$

\section{The observables scheme dependence}

\subsection{The RS invariant expressions}
Consider the case when some observable $d$ is expressed over the powers
of coupling constant  perturbatively
\begin{eqnarray}
\label{obs}d(x)=a(x)\left(1+\sum \limits_{i=1}d_ia(x)^i\right).
\end{eqnarray}
The renormalization scheme changing affects the expansion coefficients, so
does RGE and the coupling constant. The natural physical demand is the 
observable scheme independence. Initially, there was the expansion over
the scheme dependent function powers with the scheme dependent coefficients,
but expressing $a(x)$ as the two-loop function power series~(\ref{expansion})
we obtain the expansion over the scheme-independent two-loop coupling constant
\begin{eqnarray}d(x)=\widetilde{a}(x')+\left(k_2+d_1 k_1^2\right)
\widetilde{a}^2(x')+\left(k_3+2 d_1 k_1 k_2+d_2 k_1^3\right)
\widetilde{a}^3(x')+\end{eqnarray}
$$+\left(k_4+d_1 \left(2 k_1 k_3 + k_2^2\right)+
3 d_2 k_1^2 k_2+d_3 k_1^4\right) \widetilde{a}^4(x')+\,.\,.\,.,\,
x'=\frac{Q^2}{\Lambda_0^2}$$
This leads to an invariance of this expansion coefficients.
$k_2$ corresponds to the scale parameter $\Lambda$ changing
$k_2=-\ln \frac{\Lambda^2}{\Lambda_0^2}$.
Using the expressions for $k_i$ we can get invariant combinations
\begin{eqnarray}\label{rs1}
\rho_1=d_1-\ln \frac{\Lambda^2}{\Lambda_0^2}
\end{eqnarray}
\begin{eqnarray}\label{rs2}
\rho_2=\beta_2+ d_2+\left(\ln \frac{\Lambda^2}{\Lambda_0^2}\right)^2+
\left(-\beta_1+2 d_1\right) \ln \frac{\Lambda^2}{\Lambda_0^2}
\end{eqnarray}
\begin{eqnarray}\label{rs3}
\rho_3=\frac 1 2 \beta_3 +d_3+2\beta_2 d_1+
\left(-3 d_2-3 \beta_2-2\beta_2 d_1 \right) \ln \frac{\Lambda^2}{\Lambda_0^2}+
\left(3 d_1+\frac 5 2 \beta_1  \right)  \left(\ln \frac{\Lambda^2}{\Lambda_0^2}\right)^2+
\left(\ln \frac{\Lambda^2}{\Lambda_0^2}\right)^3
\end{eqnarray}
$$\,.\ \,.\ \,.\ $$
These combinations are invariant after the defining of the parameter $\Lambda_0$
certain value. The most prudent way is to make arbitrary parameter
$\Lambda_0$ close to the scale parameter physical values. Further we suppose
$\Lambda_0=\Lambda_{\overline{MS}}$. Substituting $d_1-\rho_1$ for
$\ln\frac{\Lambda^2}{\Lambda_0^2}$ we get
\begin{eqnarray}\label{inv2}
\rho_2=-d_1^2-\beta_1 d_1+\beta_2+d_2+\beta_1 \rho_1+\rho_1^2.
\end{eqnarray}
Similarly,  
$$ -d_1^2-\beta_1 d_1-\frac14\beta_1^2+\beta_2+d_2-inv.$$
$$ -d_1^2-\beta_1 d_1+\beta_2+d_2-inv.$$
The first invariant combination was obtained by Stevenson~\cite{stevenson,mat&stev},
the second one was used by Ra\c cka for introducing of the scheme dependence
criteria~\cite{raczka}. These three expression differs by scheme invariant terms. However,
the coeficients $\rho_1, \rho_2,...$ has the definite status. They are not
just invariants but the coefficients of observable expansion over the powers
of the scheme invariant two-loop coupling constant
\begin{eqnarray}\label{expansion1}
d(x)=\widetilde a(x')+\rho_1 \widetilde a^2(x')+\rho_2 \widetilde a^3(x')+...
\end{eqnarray}

\subsection{The scheme dependence estimate}
The changing of RS affects the scale parameter value according to the
following precise expression~\cite{stevenson, celmaster}. When the
new coupling constant is defined as the power series of the old one
$$a'=a+q_1a^2+...,\mbox{ then}$$
\begin{eqnarray}\label{lambda}
\Lambda'=e^{\frac{q_1}{\beta_0}}\Lambda,\, \beta_0=\frac94.
\end{eqnarray}
That is true only for the precise expressions when all perturbative terms are
summed. But the formula~(\ref{lambda}) is inapplicable for the truncated series,
its using leads to the scheme dependence augmenting. The expressions for the
$\beta$-function coefficients revaluation can be found at~\cite{bsh}.\par

Consider the three-loop case when $\beta_1, \beta_2, d_1, d_2$ are known,
we can also consider $\Lambda_0$ as defined. All multi-loop coefficients are
unknown, so we suppose them to be zero. In order to provide $\rho_1$ and
$\rho_2$ invariance $d_2$ and $d_2$ can be expressed as
$$d_1=\rho_1+\ln \frac{\Lambda^2}{\Lambda_0^2} $$
$$d_2=\rho_2-\left(\ln \frac{\Lambda^2}{\Lambda_0^2}\right)^2+
\beta_1 \ln \frac{\Lambda^2}{\Lambda_0^2}-\beta_2+
2 d_1 \ln \frac{\Lambda^2}{\Lambda_0^2}.$$
$\rho_3$ can not be invariant since we suppose
$\beta_3, d_3=0$. That is here, where we lose the observable invariance that
results from the perturbative series truncation. $\rho_3$ scheme dependence can
be considered as the measure of the total observable scheme dependence
$$\rho_3(\Lambda,\beta_2)=2\rho_1\beta_2+
\left( -2\beta_2-3\rho_2+2\rho_1\beta_1\right)\ln \frac{\Lambda^2}{\Lambda_0^2}+
\left( -3\rho_1-\frac52\beta_1\right)\left(\ln \frac{\Lambda^2}{\Lambda_0^2}\right)^2+
\left(\ln \frac{\Lambda^2}{\Lambda_0^2}\right)^3.
$$
The observable modification when changing the RS can be expressed
$$\Delta f=\Delta\rho_3\left(\ln \frac{\Lambda^2}{\Lambda_0^2},\beta_2\right)\widetilde a^4+
\Delta\rho_4\left(\ln \frac{\Lambda^2}{\Lambda_0^2},\beta_2\right)
\widetilde a^5+...\, \,\mbox{}$$
$$\Delta\rho_3=\Delta\beta_2\frac{d\rho_3}{d\beta_2}+
\Delta\left(\ln \frac{\Lambda^2}{\Lambda_0^2}\right)\frac{d\rho_3}
{d\ln \frac{\Lambda^2}{\Lambda_0^2}}.$$

\subsection{The `optimal' choice of the scale parameter value}
Let us consider again the three-loop case, when $\beta_1, \beta_2$,
$d_1, d_2$ and $\Lambda_0$ are defined.
The $\rho_1(\Lambda,d_1)$ and $\rho_2(\Lambda,d_1,d_2,\beta_2)$ invariance allows
to present $d_1$ and $d_2$ as $d_1(\Lambda)$, $d_2(\Lambda,\beta_2)$.
Four parameters $\beta_2, d_1, d_2, \ln \frac{\Lambda^2}{\Lambda_0^2}$ that
characterise the scheme at the three-loop level
are related by two equations, so we could exclude instead, for example,
$d_1$ and $\beta_2$. All numerical estimates made in this paper are for
$N_f=3$.\par

The scheme dependence of $\rho_3$ can be got over by choosing the scale
parameter. Let us require the invariance of
$\rho_3(\Lambda,d_1,d_2,\beta_2,d_3=0,\beta_3=0)$, this leads to the defining
of the scale parameter as a function of $\beta_2$. Thus, the scheme dependence
comes from $\rho_4(\beta_2)$ and it can not be eliminated at the three-loop
level.\par

The method presented in this paper provides the $\rho_3$ invariance.
The observable variation, when the current RS is modified, can be written as
$$\Delta f=(C\Delta\beta_2+O(\Delta\beta_2) )\widetilde a^5+
O(a^6),$$
\begin{eqnarray}C=\frac{d\rho_4}{d\beta_2}-\frac{d\rho_4}{d\ln \frac{\Lambda^2}{\Lambda_0^2}}
\frac{\frac{d\rho_3}{d\beta_2}}
{\frac{d\rho_3}{d\ln \frac{\Lambda^2}{\Lambda_0^2}}}=
-\frac83\beta_2+\rho_2+2\rho_1\frac{-6\rho_1-3\beta_1\beta_2+3\rho_2\beta_1
+8\rho_1\beta_2}{-2\beta_2+3\rho_2+2\rho_1\beta_1}.
\end{eqnarray}
So, at the three-loop level theoretical uncertainty can be reduced to the
$a^5$ order. This means the scheme dependence reduction comparing to the case,
when the prescription~(\ref{lambda}) is used.\par

The new criterion of the preferable RS is the minimal $C$ value. Its examination
for some observables is presented in the next section.\par

\subsection{The RS new classification}
The method suggested here let us reduce the scheme dependence. However,
it can not eliminate it, so it is interesting to make some hierarchy
of different RS using the new criterion. For this purpose we should know
$\rho_1$, $\rho_2$ and $\beta_2$ values.\par

As the first observable example we chose the Adler $D$-function, that is
defined as the logarithmic derivative of the vector current correlation
$\Pi(Q^2)$
\begin{eqnarray}
D(Q^2)=Q^2\frac{d\Pi(Q^2)}{dQ^2}=3\left(\sum_fQ_f^2\right)\left(1+d(Q^2)\right)
\mbox{, where}
\end{eqnarray}
$$
d(Q^2)=a(x)(1+d_1a(x)+d_2a^2(x)+...),\, x\equiv\frac{Q^2}{\Lambda^2}.
$$
The perturbative result at the third order (NNLO) can be found
from~\cite{gorishny&kataev&larin}. In the following table the `optimal' scale
parameter values and the new criterion values are presented.
$$\begin{array}{ccccccccc}\label{C_d}
 & d_1 & d_1^{opt} & d_2 & d_2^{opt} & b_2 & \frac{\Lambda}{\Lambda_{\overline{MS}}}& \frac{\Lambda^{opt}}{\Lambda_{\overline{MS}}}&C_d\\
\overline{MS} & 0.73 & 0.73 & 1.26 & 1.26 & 0.88 & 1 & 1 & 4.07\\ 
PMS & 0 & 0.92 & -0.52 & -0.52 & 1.55 & 0.70 & 1.1 & 2.28\\ 
ECH & 0 & 0.77 & 0 & 1.20 & 1.03 & 0.70 & 1.02 & 3.70\\ 
't H & -1.47 & 0.54 & 1.93 & 1.75 & 0 & 0.37 & 0.91 & 6.07\\ 
V & -0.048 & 2.08 & -4.17 & 1.81 & 5.17 & 0.68 & 1.96 & -15.2  
\end{array}$$
The Adler function scheme dependence was studied earlier~\cite{eidelman, chyla&kataev&larin}.
In this paper the optimal $\Lambda$ value was obtained for the 
$\overline{MS}$-scheme, for the PMS\footnote{PMS stands for the principle of
minimal sensivity}-optimyzed~\cite{stevenson} scheme,
the effective charge approach (ECH) of Grunberg~\cite{grunberg_old, kataev&kras},
the t'Hooft RS and V-scheme.\par

The examination of the $e^+e^-$-annihilation into hadrons ratio $R(s)$
(that is the mesurable quantity)
is presented in the next table
$$\begin{array}{ccccccccc}\label{C_r}
 & r_1 & r_1^{opt} & r_2 & r_2^{opt} & b_2 & \frac{\Lambda}{\Lambda_{\overline{MS}}}& \frac{\Lambda^{opt}}{\Lambda_{\overline{MS}}}&C_r\\
\overline{MS} & 0.73 & 0.73 & 1.26 & 1.26 & 0.88 & 1 & 1 & -5.81\\ 
PMS & 0.73 & 0.57 & -3.81 & -3.04 & 1.55 & 0.70 & 0.92 & -7.87\\ 
ECH & 0 & 0.68 & 0 & -3.28 & -2.27 & 0.70 & 0.976 & -6.26\\ 
't H & -1.47 & -3.90 & -1.35 & 10.45 & 0 & 0.37 & 0.094 & -118\\ 
V & -0.048 & 0.29 & -7.47 & -7.12 & 5.17 & 0.68 & 0.80 & -18.6  
\end{array}$$

The method can be easily generalised onto other observables such as
the $\tau$-decay ratio, the Bjorken sum rule etc.\par

The new criterion values obtained above can not serve for a definite choice
of the preferable RS. Thus, the PMS is the optimal RS for the Adler-function
among the
RS set considered, but the $R(x)$-function lowest criterion absolute value is
for the $\overline{MS}$-scheme. However, $\overline{MS}$, PMS and ECH have approximately
the same criterion values for both cases and other two schemes have very
large values. We can conclude that the new criterion $C$ can not be
considered as a universal value characterising the RS but it is
suitable for `bad' schemes sorting out.

\section{Conclusions}
The method suggested in this paper allows to express the
renorm-group equation solutions over the powers of the two-loop one. This can
be very useful for some numerical estimates, especially after the
analyticization procedure~\cite{dv96, dv97, dv98} applying.\par

This method helps to reduce the scheme dependence, to analyse the problem better
and get some estimates of this dependence value. The {\bf only} value that is
adjusted is the scale parameter $\Lambda$. We can conclude that the scheme
dependence can be minimised just by the `optimal' revaluation of the
experimentally extracted value $\Lambda$ without any additional
requirements.\par

The idea about the observable expansion over the two-loop coupling constant is
interesting to compare with the observable expansion in the recently
published by C.~Maxwell~\cite{maxwell} paper. However, we used another
way to obtain this expansion, results obtained in this paper are 
different too. \par

\bigskip %

{\large\bf Acknowledgements}
\vspace{0.1cm}

I would like to thank D.V.~Shirkov and I.L.~Solovtsov for their
invaluable help in the creation of this method. The author also thanks
A.L.~Kataev for very useful discussion and for constructive criticism of
the original version of this paper.

\end{document}